\documentclass{PoS}

\title{Status of the Unitarity Triangle Analysis}

\ShortTitle{UTA Status}

\author{M.~Bona, M.~Pierini\\
       CERN, CH-1211 Geneva 23, Switzerland}
\author{M.~Ciuchini\\
        INFN,  Sezione di Roma Tre, I-00146 Roma, Italy}
\author{E.~Franco, L.~Silvestrini\\
        INFN, Sezione di Roma, I-00185 Roma, Italy}
\author{V.~Lubicz, \speaker{Cecilia Tarantino}\\
        INFN,  Sezione di Roma Tre, I-00146 Roma, Italy\\
        Dipartimento di Fisica, Universit{\`a} di Roma Tre,
        I-00146 Roma, Italy} 
\author{G.~Martinelli\\
        INFN, Sezione di Roma, I-00185 Roma, Italy\\
        Dipartimento di Fisica, Universit\`a di Roma ``La
        Sapienza'', I-00185 Roma, Italy} 
\author{F.~Parodi, C.~Schiavi\\
       Dipartimento di Fisica, Universit\`a di Genova and INFN, I-16146
       Genova, Italy} 
\author{V.~Sordini\\
       ETH Zurich, HG Raemistrasse 101, 8092 Zurich, Switzerland}
\author{A.~Stocchi\\
       Laboratoire de l'Acc\'el\'erateur Lin\'eaire, IN2P3-CNRS et
       Universit\'e de Paris-Sud, BP 34, 
       F-91898 Orsay Cedex, France}
\author{V.~Vagnoni\\
        INFN, Sezione di Bologna,  I-40126 Bologna, Italy}

\abstract{We present an update of the Unitarity Triangle (UT) analysis, within the Standard Model (SM) and beyond.
Within the SM the main novelties are the inclusion in $\varepsilon_K$ of the contributions of $\xi$ and $\phi_{\varepsilon} \neq \pi/4$ pointed out by A.~J.~Buras and D.~Guadagnoli, and an accurate prediction of BR$(B \to \tau \nu)$, by using the indirect determination of $|V_{ub}|$ from the UT fit, which can be compared to the present experimental result.
In the generalization of the UT analysis to investigate New Physics (NP) effects, the estimate of $\xi$ is more delicate and only the effect of $\phi_{\varepsilon} \neq \pi/4$ has been included. We confirm an hint of NP in the $B_s$-$\bar B_s$ mixing  at the $2.9 \sigma$ level, which makes a comparison with new experimental data certainly desired. 
}

\FullConference{European Physical Society Europhysics Conference on High Energy Physics,
EPS-HEP 2009,\\
		 July 16 - 22 2009\\
		 Krakow, Poland}

\begin{document}

\begin{figure}[tb]	
    \center{\includegraphics[width=0.5\textwidth]{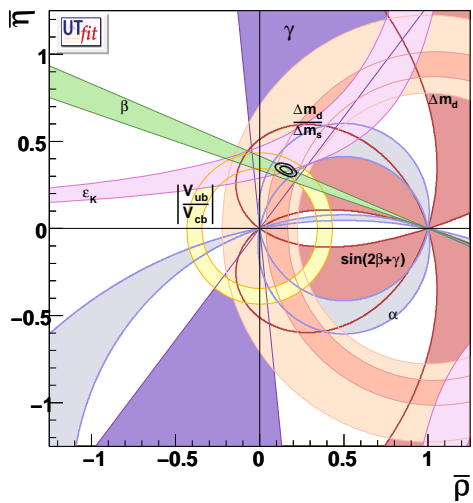}} 
    \caption{Result of the UT fit within the SM. The contours display the 68\% and 95\% probability regions selected by the
   fit in the $(\bar \rho, \bar \eta)$-plane. The 95\% probability regions selected by the single constraints are also shown.}
\label{fig:all}
\end{figure}

We present an update of the Unitarity Triangle (UT) analysis performed by the UTfit collaboration following the method described in refs.~\cite{Ciuchini:2000de,Bona:2005vz}.
Within the Standard Model (SM), we have included in $\varepsilon_K$ the contributions of $\xi$ and $\phi_\varepsilon \neq \pi/4$ which, as pointed out in~\cite{Buras:2008nn}, decrease the SM prediction for $\varepsilon_K$ by $\sim 8$\%.
We observe, as main result of the UT analysis, that the CKM matrix turns out to be consistently overconstraint
and the CKM parameters $\bar \rho$ and $\bar \eta$ are accurately determined: $\bar \rho=0.154\pm0.021$, $\bar \eta=0.340\pm0.013$~\cite{UTfitwebpage}.
The UT analysis has thus established that the CKM matrix is the dominant source of flavour mixing and CP-violation and that New Physics (NP) effects can at most represent a small correction to this picture.
We note, however, that the new contributions in $\varepsilon_K$ generate some tension in particular between the constraints provided by the experimental measurements of $\varepsilon_K$ and $\sin 2 \beta$ (see fig.~\ref{fig:all}).
As a consequence, the indirect determination of $\sin 2 \beta$ turns out to be larger than the experimental value by $\sim 2.0 \sigma$.\footnote{For an alternative indirect determination of $\sin 2 \beta$ which does not rely and is thus free from the hadronic uncertainty in $|V_{ub}|$, see ref.~\cite{Lunghi:2008aa}.}
We observe that since new unquenched results for the bag-parameter $B_K$ tend to lie below the older quenched results~\cite{lattice09}, an update of the input value for $B_K$, which is in program, is expected to enhance this $\varepsilon_K$-$\sin 2 \beta$ tension.

Recently, we have shown~\cite{Bona:2009cj} how to use the UT fit to improve the prediction of BR$(B \to \tau \nu)$ in the SM, thanks to a better determination of $|V_{ub}|$ and $f_B$. Within the SM the UT fit prediction for BR$(B \to \tau \nu)$ is found to deviate from the experimental measurement~\cite{HFAG} by $\sim 2.5 \sigma$. Even allowing for NP effects in $\Delta F=2$ processes, while assuming negligible NP contributions to the $B \to \tau \nu$ decay amplitude, a $\sim 2.2 \sigma$ deviation from the experimental value is found.

We now present the update of the NP UT analysis, that is the UT analysis generalized to include possible NP effects.
In $\varepsilon_K$ we have taken into account the effect of $\phi_{\varepsilon} \neq \pi/4$, while the $\xi$ contribution, which beyond minimal flavour violation (MFV)~\cite{D'Ambrosio:2002ex,Buras:2000dm} is affected  by a large uncertainty~\cite{Buras:2009pj}, is not included. 
This analysis consists first in generalizing the relations among the experimental observables and the elements of the CKM matrix, introducing effective model-independent parameters that quantify the deviation of the experimental results from the SM expectations.
The possible NP effects considered in the analysis are those entering neutral meson mixing.
Thanks to recent experimental developments, in fact, these $\Delta F=2$ processes turn out to provide stringent constraints on possible NP contributions.
In the case of $B_{d,s}$-$\bar B_{d,s}$ mixing, a complex effective parameter is introduced, defined as
\begin{equation}
\qquad \qquad \qquad \qquad C_{B_{d,s}}\,e^{2 i \phi_{B_{d,s}}} = \frac{\langle B_{d,s} | H_{eff}^{full}| \bar B_{d,s} \rangle}{\langle B_{d,s} | H_{eff}^{SM}| \bar B_{d,s} \rangle}\,,
\end{equation}
being $H_{eff}^{SM}$ the SM $\Delta F=2$ effective Hamiltonian and $H_{eff}^{full}$ its extension in a general NP model, and with $C_{B_{d,s}}=1$ and $\phi_{B_{d,s}}=0$ within the SM.
All the mixing observables are then expressed as a function of these parameters and the SM ones (see refs.~\cite{Bona:2005eu,Bona:2006sa,Bona:2007vi} for details).
In a similar way, for the  $K$-$\bar K$ system, one can write
\begin{equation}
\qquad C_{\epsilon_K} = \frac{Im[\langle K | H_{eff}^{full}| \bar K \rangle]}{Im[\langle K | H_{eff}^{SM}| \bar K \rangle]}\,,\qquad \qquad
C_{\Delta m_K} = \frac{Re[\langle K | H_{eff}^{full}| \bar K \rangle]}{Re[\langle K | H_{eff}^{SM}| \bar K \rangle]}\,,
\end{equation}
with $C_{\epsilon_K}=C_{\Delta m_K}=1$ within the SM.

In this way, the combined fit of all the experimental observables selects a region of the $(\bar \rho, \bar \eta)$ plane ($\bar \rho=0.177\pm0.044$, $\bar \eta=0.360\pm0.031$) which is consistent with the results of the SM analysis,
and it also constraints the effective NP parameters.

For $K$-$\bar K$ mixing, the NP parameters are found in agreement with the SM expectations. In the $B_d$ system, the mixing phase $\phi_{B_d}$ is found $\simeq 1.5 \sigma$ away from the SM expectation, reflecting a slight tension between the direct measurement of $\sin 2 \beta$ and its indirect determination from the other UT constraints.

\begin{figure}[tb]	
    \center{\includegraphics[width=0.5\textwidth]{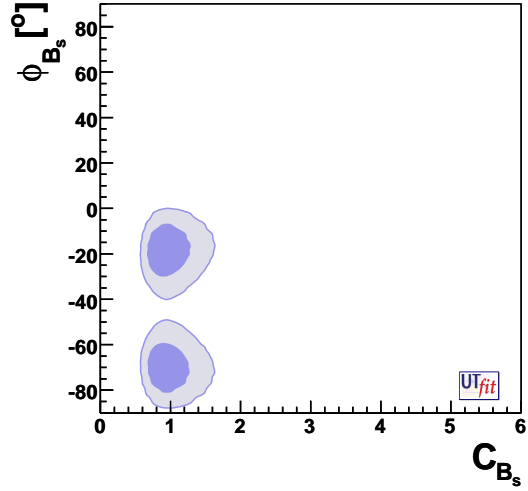}} 
    \caption{68\% (dark) and 95\% (light) probability regions in the ($C_{B_s},\phi_{B_s}$)-plane.}
\label{fig:phiBs}
\end{figure}

The $B_s$-meson sector, where the tiny SM mixing phase $\sin 2 \beta_s \simeq 0.041(4)$ could be highly sensitive to a NP contribution, represents a privileged environment to search for NP.
In this sector, an important experimental progress has been achieved at the Tevatron collider in 2008 when both the CDF~\cite{Aaltonen:2007he} and D0~\cite{:2008fj} collaborations published the two-dimensional likelihood ratio for the width difference $\Delta \Gamma_s$ and the phase $\phi_s=2(\beta_s-\phi_{B_s})$, from the tagged time-dependent angular analysis of the decay $B_s \to J_{\psi} \phi$. 
Updating the UTfit analysis of ref.~\cite{Bona:2008jn}, by combining the CDF and D0 results including the now available D0 two-dimensional likelihood without assumptions on the strong phases, we find $\phi_{B_s}=(-69\pm7)^\circ \cup (-19\pm8)^\circ$, which is $2.9\sigma$ away from the SM expectation $\phi_{B_s}=0$ (see fig.~\ref{fig:phiBs}). A deviation of more than $2 \sigma$ is found also by the Heavy Flavour Averaging Group (HFAG)~\cite{HFAG} ($2.2 \sigma$) and by CKMfitter~\cite{Deschamps:2008de} ($2.5 \sigma$), by combining the Tevatron results with some differences in the statistical approach.

It will be interesting to see if this hint of NP will be confirmed once the Tevatron measurements will improve, in particular when the CDF collaboration will make the
new likelihood, based on an enlarged data sample of $2.8 fb^{-1}$, publicly available.
We note that this NP signal would be not only a signal of physics beyond the SM but more in general beyond MFV, since a value of $\phi_{B_s}$ different from zero can only be an effect of a new source of flavour violation different from the Yukawa couplings.


\begin{thebibliography}{99}
\bibitem{Ciuchini:2000de}
  M.~Ciuchini {\it et al.},
  JHEP {\bf 0107} (2001) 013
  [hep-ph/0012308].

\bibitem{Bona:2005vz}
  M.~Bona {\it et al.}  [UTfit Collaboration],
  JHEP {\bf 0507} (2005) 028
  [hep-ph/0501199].

\bibitem{Buras:2008nn}
  A.~J.~Buras and D.~Guadagnoli,
  Phys.\ Rev.\  D {\bf 78} (2008) 033005
  [0805.3887 [hep-ph]].


\bibitem{UTfitwebpage}
The UTfit Collaboration, http://www.utfit.org/. 

\bibitem{Lunghi:2008aa}
  E.~Lunghi and A.~Soni,
  Phys.\ Lett.\  B {\bf 666} (2008) 162
  [arXiv:0803.4340 [hep-ph]].

\bibitem{lattice09}
V.~Lubicz, talk at LATTICE'09, July 26th - 31th, Beijing. 

\bibitem{Bona:2009cj}
  :.~M.~Bona {\it et al.}  [UTfit Collaboration],
  arXiv:0908.3470 [hep-ph].

\bibitem{HFAG}
The Heavy Flavour Averaging Group (HFAG), http://www.slac.stanford.edu/xorg/hfag/.

\bibitem{D'Ambrosio:2002ex}
  G.~D'Ambrosio {\it et al.},
  Nucl.\ Phys.\  B {\bf 645} (2002) 155
  [hep-ph/0207036].

\bibitem{Buras:2000dm}
  A.~J.~Buras {\it et al.},
  Phys.\ Lett.\  B {\bf 500} (2001) 161
  [hep-ph/0007085].

\bibitem{Buras:2009pj}
  A.~J.~Buras and D.~Guadagnoli,
  Phys.\ Rev.\  D {\bf 79} (2009) 053010
  [arXiv:0901.2056 [hep-ph]].


\bibitem{Bona:2005eu}
  M.~Bona {\it et al.}  [UTfit Collaboration],
  JHEP {\bf 0603} (2006) 080
  [hep-ph/0509219].

\bibitem{Bona:2006sa}
  M.~Bona {\it et al.}  [UTfit Collaboration],
  Phys.\ Rev.\ Lett.\  {\bf 97} (2006) 151803
  [hep-ph/0605213].

\bibitem{Bona:2007vi}
  M.~Bona {\it et al.}  [UTfit Collaboration],
  JHEP {\bf 0803} (2008) 049
  [0707.0636 [hep-ph]].

\bibitem{Aaltonen:2007he}
  T.~Aaltonen {\it et al.}  [CDF Collaboration],
  Phys.\ Rev.\ Lett.\  {\bf 100} (2008) 161802
  [0712.2397 [hep-ex]].

\bibitem{:2008fj}
  V.~M.~Abazov {\it et al.}  [D0 Collaboration],
  Phys.\ Rev.\ Lett.\  {\bf 101} (2008) 241801
  [0802.2255 [hep-ex]].

\bibitem{Bona:2008jn}
  M.~Bona {\it et al.}  [UTfit Collaboration],
  0803.0659 [hep-ph].

\bibitem{Deschamps:2008de}
  O.~Deschamps,
  arXiv:0810.3139 [hep-ph].

\end{thebibliography}
\end{document}